\documentclass[a4paper,twoside,final,12pt,titlepage]{article}
\usepackage{subfigure}
\usepackage{amsmath}
\usepackage{amsfonts}
\usepackage{amssymb}
\usepackage{graphicx}%
\usepackage{eufrak}
\providecommand{\U}[1]{\protect\rule{.1in}{.1in}}

\newskip\humongous \humongous=0pt plus 1000pt minus 1000pt

\newif\ifdtup

\catcode`\@=11
\@addtoreset{equation}{section}

\def\@normalsize{\@setsize\normalsize{15pt}\xiipt\@xiipt
\abovedisplayskip 14pt plus3pt minus3pt\belowdisplayskip \abovedisplayskip
\abovedisplayshortskip \z@ plus3pt\belowdisplayshortskip 7pt plus3.5pt minus0pt}
\def\small{\@setsize\small{13.6pt}\xipt\@xipt
\abovedisplayskip 13pt plus3pt minus3pt\belowdisplayskip \abovedisplayskip
\abovedisplayshortskip \z@ plus3pt\belowdisplayshortskip 7pt plus3.5pt minus0pt
\def\@listi{\parsep 4.5pt plus 2pt minus 1pt
\itemsep \parsep
\topsep 9pt plus 3pt minus 3pt}}
\relax
\catcode`@=12
\catcode`\@=11
\begin{document}
\begin{titlepage}
\renewcommand{\thefootnote}{\fnsymbol{footnote}}
\bigskip
\bigskip
\bigskip
\bigskip
\bigskip
\begin{center}
{\Large  {\bf Soliton Junctions \\ } } {\Large  {\bf  in the Large
Magnetic Flux Limit} }
\end{center}
\renewcommand{\thefootnote}{\fnsymbol{footnote}}
\bigskip
\begin{center}
{\large   Stefano {\sc Bolognesi}\footnote{bolognesi@nbi.dk}} and
{\large Sven
Bjarke {\sc Gudnason}\footnote{gudnason@nbi.dk}}
\end{center}
\begin{center}
{\it      \footnotesize The Niels Bohr Institute, Blegdamsvej 17,
DK-2100 Copenhagen \O, Denmark  } \vskip 0.15cm
\end {center}
\renewcommand{\thefootnote}{\arabic{footnote}}
\setcounter{footnote}{0}
\bigskip
\bigskip
\bigskip
\noindent
\begin{center} {\bf Abstract} \end{center}
We study the flux tube junctions in the limit of large magnetic
flux. In this limit the flux tube becomes a {\it wall vortex} which
is a wall of negligible thickness (compared to the radius of the
tube) compactified on a cylinder and stabilized by the flux inside.
This wall surface can also assume different shapes that correspond
to soliton junctions. We can have a flux tube that ends on a wall, a
flux tube that ends on a monopole and more generic configurations
containing all three of them. In this paper we find the differential
equations that describe the shape of the {\it wall vortex} surface
for these junctions. We will restrict to the cases of cylindrical
symmetry. We also solve numerically these differential equations for
various kinds of junctions. We finally find an interesting relation
between soliton junctions and dynamical systems. \vfill
\begin{flushleft}
June, 2006
\end{flushleft}
\end{titlepage}

\section{Introduction}

In a recent series of works \cite{wallvortex,wallvortexdue,proof} we studied
the behavior of the Abrikosov-Nielsen-Olesen (ANO) vortex \cite{Abrikosov} in
the large $n$ limit, where $n$ is the number of quanta of magnetic flux
carried by the vortex. We have seen that in this limit the ANO vortex becomes
essentially a bag, analogous to the bag models of hadrons \cite{MIT,SLAC}. We
now briefly review the essential results.

\begin{figure}[h!tb]
\begin{center}
\includegraphics[
height=2.1811in,
width=4.0542in
]{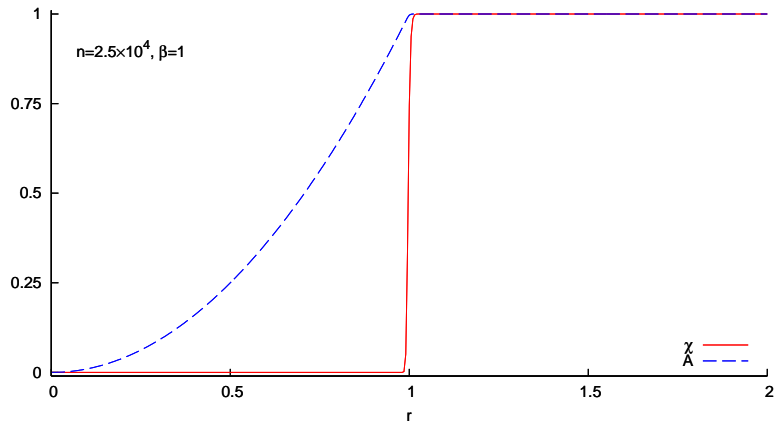}
\end{center}
\caption{{\protect\footnotesize {This plot is one of the outcomes of
      the numerical 
analysis made in \cite{proof}. It shows the profile functions $q(r)$
(red/solid line) and $A(r)$ (blue/dashed line) for the BPS potential
($\beta=1$) and winding number $n=25,000$. }}}%
\label{proof}%
\end{figure}

The theory under consideration is the Abelian-Higgs model
\begin{equation}
\mathcal{L}=-\frac{1}{4}F_{\mu\nu}F^{\mu\nu}-|(\partial_{\mu}-ieA_{\mu}%
)q|^{2}-V(|q|)\ , \label{BasicVortex}%
\end{equation}
where the potential is chosen such that it has a minimum in the Higgs phase
$|q|=q_{0}\neq0$. The ANO vortex is a string-like soliton that extends in time
and one spatial dimension. The simplest way to describe it is to introduce
cylindrical coordinates $(z,r,\theta)$ and orientate it in the $\hat{z}$
direction. The fields can then be put into the following form
\begin{align}
q  &  =q_{0}e^{in\theta}\,q(r)\ ,\label{vortex}\\
A_{\theta}  &  =\frac{n}{er}\,A(r)\ .\nonumber
\end{align}
The problem is now to evaluate the profile functions $q(r)$ and $A(r)$
subjected to the boundary conditions $q(0)=0$, $q(\infty)=q_{0}$ and $A(0)=0$,
$A(\infty)=1$. The claim is that, for every Higgs-like potential $V$, in the
large $n$ limit the profiles become
\begin{align}
\lim_{n\rightarrow\infty}q(r)&\rightarrow\theta_{H}(r-R_{\mathrm{V}%
})\ ,\label{reformulation}\\
\lim_{n\rightarrow\infty}A(r)&\rightarrow\left\{
\begin{array}
[c]{cc}%
r^{2}/{R_{\mathrm{V}}}^{2} & 0\leq r\leq R_{\mathrm{V}}\ ,\\
1 & r>R_{\mathrm{V}}\ .
\end{array}
\right. \nonumber
\end{align}
This conjecture has been proved in \cite{proof} by numerical
computations. We show in Figure \ref{proof} the result obtained for
the BPS potential and $n=25,000$. The step function of the profile $q(r)$ reveals the presence of a
substructure: a domain wall interpolating between the Coulomb phase $q=0$ and
the Higgs phase $q=q_{0}$. For this reason we have named this object
wall vortex. 

The wall vortex is essentially a bag, such as the ones used in the context of
the bag models of hadrons. A domain wall of thickness $\Delta_{\mathrm{W}}$
and tension $T_{\mathrm{W}}$ is wrapped onto a cylinder of radius $R$. The bag
model is a good approximation when the thickness of the wall is small
compared to the radius of the vortex. There are three energy terms that
must be considered. The first one comes from the tension of the wall and is
proportional to the radius $R$. The second comes from the energy density of
the interior of the bag and is proportional to $R^{2}$. The third energy term
comes from the magnetic flux and is proportional to $1/R^{2}$. When they are
summed we obtain the tension as function of the radius $R$:
\begin{equation}
T(R)=\frac{2\pi n^{2}}{e^{2}R^{2}}+T_{\mathrm{W}}2\pi R+\varepsilon_{0}\pi
R^{2}\ . \label{unavariabile}%
\end{equation}
The radius of the vortex $R_{\mathrm{V}}$ is the one that minimizes this
expression. Physically we can understand it in this way. When we derive
(\ref{unavariabile}) with respect to $R$ we obtain the various forces that act
on the surface of the bag. The first two terms bring a collapse force that
tends to squeeze the tube. The third term acts instead as a pressure that
tends to expand the tube. When the two opposite forces are equilibrated we
have a stable configuration.

The aim of the present paper is to study a more generic embedding of the wall
vortex in the three dimensional space. The configuration (\ref{unavariabile})
has the maximal number of possible symmetries, it is invariant under
cylindrical rotation and $z$ translation. Now we want to relax the last
condition, namely we consider a wall that has only cylindrical symmetry. The
radius $R$ is no longer a constant but a function $R=f(z)$, and the force
balance will not be an algebraic equation but a system of differential
equations.

Solitons in the Higgs phase have received great attention in the
last years. A lot of different models in different limits have been
investigated and the jungle of solitons and soliton junctions is
enormously vast. Here we just mention the works that have mostly
influenced the present paper: solitons in nonlinear sigma models
\cite{Gauntlettuno,Gauntlettdue}, vortices and walls in
supersymmetric models with a Fayet-Ilioupoulos term
\cite{vwuno,vwdue,vwtre}, the moduli matrix approach
\cite{tokyo,tokyoo,tokyooo}, nonabelian vortices and their junction
with nonabelian monopoles \cite{HananyTong,nostrobig,iozazejarah},
the vortex-monopole-vortex junction
\cite{tongmonopole,Shifmanmonopole} and also the MQCD realization of
soliton junctions \cite{mqcd,HSZmqcd,ioejarahmqcd}. We find
particularly amazing the various relations between the two
fundamental solitons in the Higgs phase: the vortex and the wall.
More recent works are
\cite{Tongduality,Shifmanduality,Sakaiduality}. The large magnetic
flux limit is a very general approach, it applies to the
Abelian-Higgs model (\ref{BasicVortex}) that is the basic building
block of all the more sophisticated theories that contain solitons
in the Higgs phase. Our result can thus be applied also to these
theories and maybe explain some of the various relations between
walls and vortices. Some of these recent developments on solitons
are summarized in the reviews \cite{revtong,revtokyo}.

The paper is divided into two main parts. In Section
\ref{differentialequation} we derive the system of differential equations that
governs the vortex junctions.\ In Section \ref{numericalsolutions} we
explicitly study the physical solutions of the differential equations. We
conclude in \ref{conclusion}\ with comments and possible applications.

\section{The Differential Equations \label{differentialequation}}

To obtain the differential equations for the profile $r=f(z)$ we proceed in
three steps. First in Subsection \ref{redone} we redo the wall vortex analysis
interpreting the minimization of $T(R)$ (\ref{unavariabile}) as a balance of
forces acting on the wall. Then in Subsection \ref{mechanical} we find the
mechanical forces (the ones coming from the tension $T_{\mathrm{W}}$ and the
energy density $\varepsilon_{0}$) for a generic profile with cylindrical
symmetry. Finally in Subsection \ref{master} we obtain the master equation
governing the profile $f(z)$.

\subsection{Wall vortex redone \label{redone}}

The wall vortex radius is obtained by the minimization of the function $T(R).$
The derivative of (\ref{unavariabile}) divided by the perimeter $2\pi R$ gives%
\begin{equation}
-\frac{2n^{2}}{e^{2}R^{4}}+\frac{T_{\mathrm{W}}}{R}+\varepsilon_{0}=0\ .
\label{forcessimple}%
\end{equation}
This equation can be interpreted as a balance of forces per unit of area
acting on the surface of the wall. Preceding from the right to left of
(\ref{forcessimple}) we have:

\begin{itemize}
\item The energy density $\varepsilon_{0}$ is a force per unit area
  directed inwards;

\item The tension of the wall $T_{\mathrm{W}}$ divided by the radius of
curvature $R$ is a force per unit area directed inwards. Note that the other
radius of curvature of the surface is infinite and does not contribute any
extra force;

\item The remaining term $-\frac{2n^{2}}{e^{2}R^{4}}$ must be interpreted as a
force due to the magnetic field on the boundary of the cylinder and directed
outwards. To check the consistency, note that since the flux carried by the
wall vortex is $\Phi_{B}=\frac{2\pi n}{e}$ and the magnetic field is
$B=\frac{2n}{eR^{2}}$, the term $\frac{2n^{2}}{e^{2}R^{4}}$ is exactly equal
to the magnetic field energy density $\frac{B^{2}}{2}$.
\end{itemize}

\bigskip There is a simple way to understand the origin of the magnetic field
force $\frac{B^{2}}{2}$. \begin{figure}[h!tb]
\begin{center}
\includegraphics[
height=2.258in,
width=3.2188in
]{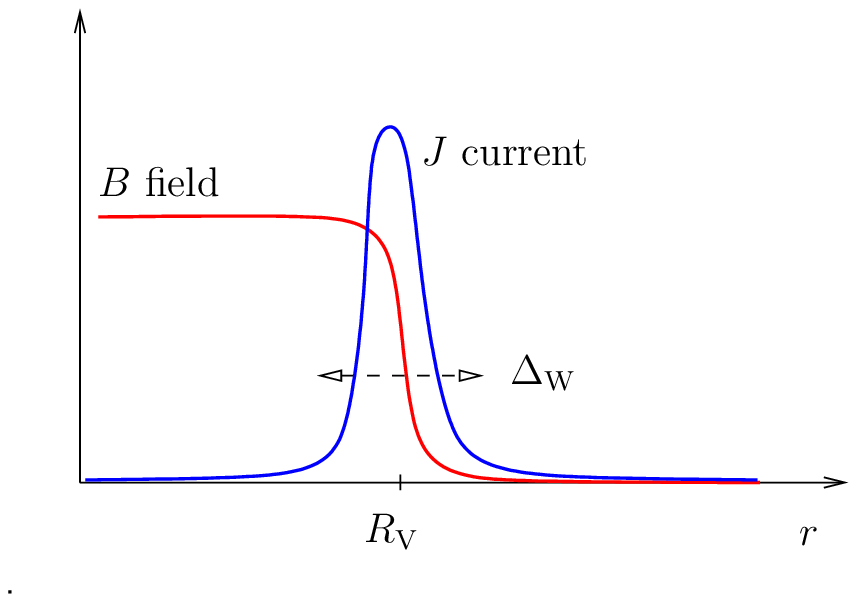}
\end{center}
\caption{{\protect\footnotesize {The magnetic field $B$ and the magnetic
current $J$ at the boundary of the wall vortex.}}}%
\label{force}%
\end{figure}We can write the magnetic field as a function of the radius $B(r)$
where $B=\frac{2n}{eR^{2}}$ inside the wall vortex, $B=0$ outside and the
transition is encoded in the function $B(r)$. Using the Maxwell equation
$\vec{\nabla}\wedge\vec{B}\ =\vec{J}\ $\ we find $J(r)=\frac{dB(r)}{dr}$. The
force per unit volume acting on density current is $\vec{F}=\vec{J}\wedge
\vec{B}$ thus the force per unit of area acting on the surface of the wall is
\begin{equation}
\int B(r)\frac{dB(r)}{dr}dr=\int d\left(  \frac{B\left(  r\right)  ^{2}}%
{2}\right)  =\left.  \frac{B^{2}}{2}\right\vert _{\mathrm{inside}} \ ,%
\end{equation}
and is directed outwards. Note that this force is independent on the precise
behavior of the function $B(r)$.

\subsection{Mechanical Forces \label{mechanical}}

Now we want to find the mechanical forces in the case of a generic profile
$r=f(z)$. We proceed in this way. First we write the mechanical energy as an
integral with respect to $z$ of the surface element times the tension
$T_{\mathrm{W}}$ plus the volume element times the energy density
$\varepsilon_{0}$:%
\begin{equation}
E_{\mathrm{mech}}=\int dz\left(  2\,\pi T_{\mathrm{W}}\,f\,\sqrt
{1+f^{\prime\,2}}\,+\pi\varepsilon_{0~}f^{2}\right)  ~.
\end{equation}
In absence of the magnetic field, the profile $f(z)$ is the solution to the
Euler-Lagrange equations obtained by varying $E_{\mathrm{mech}}$:
\begin{equation}
2\pi T_{\mathrm{W}}~\frac{1+f^{\prime\,2}-f^{\prime\prime}\,f}{(1+{f^{\prime}%
}^{2})^{3/2}}+2\pi\varepsilon_{0}~f\ =0~.
\end{equation}
Dividing by the perimeter $2\pi f$ one obtains the forces per unit area
\begin{equation}
T_{\mathrm{W}}\frac{1}{f(1+f^{\prime\,2})^{1/2}}-T_{\mathrm{W}}\frac
{~f^{\prime\prime}}{(1+{f^{\prime}}^{2})^{3/2}}+\varepsilon_{0}=0
\label{forces}%
\end{equation}
$\varepsilon_{0}$ is a force per unit area due to the internal
energy density and it does not depend on the profile. The other two
terms can be identified with the tension divided by the two radia of
curvature of the surface. To verify it we now compute the two radia
of curvature geometrically.\begin{figure}[h!tb]
\begin{center}
\includegraphics[
height=2.2105in,
width=3.7905in
]{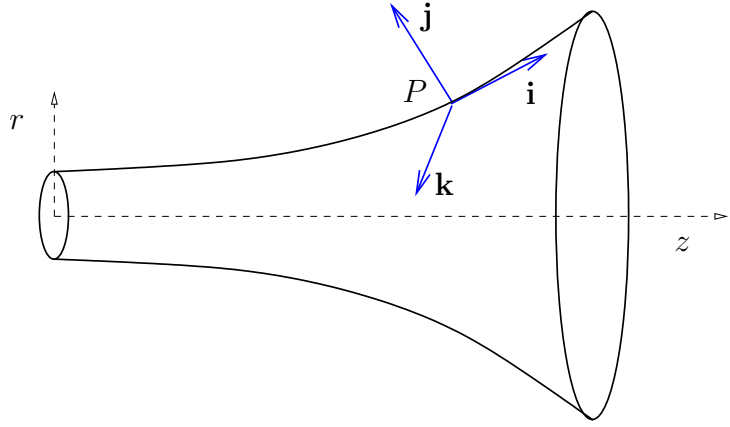}
\end{center}
\caption{{\protect\footnotesize {Geometry of the wall vortex surface.}}}%
\label{mech}%
\end{figure}
Take a point $P=(r,\phi,z)=(f(z),0,z)$, or in the Cartesian
coordinates, $(x,y,z)=(f(z),0,z)$ on the surface. The tangent vector in that
point is
\begin{equation}
\mathbf{i}=(\frac{f^{\prime}}{\sqrt{1+f^{\prime\,2}}},0,\frac{1}%
{\sqrt{1+f^{\prime\,2}}})~. \label{tangent}%
\end{equation}
The plane orthogonal to $\mathbf{i}$ and passing through $P$ is%
\begin{equation}
f^{\prime}\,(X-f)+(Z-z)=0~. \label{plan}%
\end{equation}
The unit orthonormal vectors lying on the plane (\ref{plan}) and orthogonal to
$\mathbf{i}$ are $(\mathbf{j},\,\mathbf{k})$, where
\begin{equation}
\mathbf{j}=(\frac{-1}{\sqrt{1+f^{\prime\,2}}},0,\frac{f^{\prime}}%
{\sqrt{1+f^{\prime\,2}}})~,\qquad\mathbf{k}=\mathbf{i}\wedge\mathbf{j}%
=(0,1,0)~.
\end{equation}
One radius of curvature is in the plane $(\mathbf{i},\,\mathbf{j})$ and is the
easiest one to compute
\begin{equation}
\frac{1}{R_{1}}=-\frac{f^{\prime\prime}}{(1+f^{\prime2})^{2/3}}~.
\end{equation}
It is directed outwards in case of positive $f^{\prime\prime}$ or inwards in
case of negative $f^{\prime\prime}$. The other radius of curvature is in the
plane $(\mathbf{j},\,\mathbf{k})$ defined by Eq. (\ref{plan}). The curvature
inwards that lies in the plane $(\mathbf{j},\,\mathbf{k})$ is
\begin{equation}
\frac{1}{R_{2}}=\frac{1}{f\,\sqrt{1+f^{\prime2}}}~.
\end{equation}
In summary, the force inwards and outwards at the point $P=(f(z),z)$
is respectively
\begin{equation}
F_{1}=\frac{1}{f\,\sqrt{1+f^{\prime2}}}\,T_{\mathrm{W}};\qquad F_{2}%
=-\frac{f^{\prime\prime}}{(1+f^{\prime2})^{2/3}}\,T_{\mathrm{W}}~,
\end{equation}
in agreement with Eq. (\ref{forces}).

\subsection{\bigskip Master equations \label{master}}

Now we finally write the differential equations that define the wall vortex
profile $r=f(z)$. First of all we have the Maxwell equations $\vec{\nabla
}\cdot\vec{B}=0$ and $\vec{\nabla}\wedge\vec{B}=0$ inside the profile. We can
rewrite them introducing the magnetic scalar potential $\varphi$
defined by $\vec{B}=\vec{\nabla}\varphi$. In this way the second
Maxwell equation is 
automatically satisfied while the first becomes the Laplace equation for the
magnetic scalar potential. The boundary conditions are that the $\vec{B}$
field is tangent to the surface of the wall and parallel to the vector
$\mathbf{i}$ defined in\ Eq. (\ref{tangent}). In this way the magnetic flux is
constant along $z$ and is equal to $\frac{2\pi n}{e}$. In summary
\begin{equation}
\Delta\varphi=0~,\qquad\ \ \ \vec{\nabla}\varphi\parallel\mathbf{i~,\qquad
\ \ }\Phi_{B}=\frac{2\pi n}{e}~. \label{uno}%
\end{equation}
The other equation is the balance of the forces acting on the wall, that is
the generalization of (\ref{forcessimple}) in the case of generic
cylindrically symmetric surface:%
\begin{equation}
-\left.  \frac{B^{2}}{2}\right\vert _{\mathrm{wall}}+T_{\mathrm{W}}%
\frac{1+f^{\prime2}-~f^{\prime\prime}f}{f(1+f^{\prime2})^{3/2}}+\varepsilon
_{0}=0~, \label{due}%
\end{equation}
where by $\left.  B^{2}/2\right\vert _{\mathrm{wall}}$ we mean the magnetic
field force evaluated at the wall surface. Equations (\ref{uno}) and
(\ref{due}) are a system of coupled differential equations (one partial and
one ordinary) that defines the profile $f(z)$.

\section{The Junctions \label{numericalsolutions}}

In this section we will discuss the physical solutions to the master equation
(the system (\ref{uno}) and (\ref{due})). We first make a rescaling in order
to simplify the master equation. We rescale the lengths of a quantity
$l\rightarrow R_{\mathrm{V}}l$ such that the radius of the vortex now is $1$.
Note that this rescaling applies to all the lengths, so $f\rightarrow
R_{\mathrm{V}}f$, $f^{\prime}\rightarrow f^{\prime}$ and $f^{\prime\prime
}\rightarrow R_{\mathrm{V}}^{-1}f^{\prime\prime}.$ We rescale also the
magnetic field such that $B=1$ in the case of the wall vortex. Thus
the rescaled master equations are:
\begin{equation}
\Delta\varphi=0~,\qquad\vec{\nabla}\varphi\parallel\mathbf{i~,\qquad}\Phi
_{B}=\pi\mathbf{~;} \label{Laplace}%
\end{equation}%
\begin{equation}
-\left.  B^{2}\right\vert _{\mathrm{wall}}+\rho\frac{1+f^{\prime2}%
-~f^{\prime\prime}f}{f(1+f^{\prime2})^{3/2}}+(1-\rho)=0~. \label{rescaled}%
\end{equation}
It depends only on one parameter $\rho$ defined by%
\begin{equation}
\frac{\rho}{1-\rho}=\frac{T_{\mathrm{W}}}{R_{\mathrm{V}}\varepsilon_{0}}~.
\end{equation}
The parameter $\rho$ can take on values from $0$ to $1$. If $\rho=0$ the
collapse force is due only to the internal energy density (MIT bag), while if
$\rho=1$ it is due only to the tension of the wall (SLAC bag).

\subsection{\textquotedblleft Near vortex\textquotedblright\ approximation
\label{nearvortexapproximation}}

\begin{figure}[h!tb]
\begin{center}
\includegraphics[
height=2.2736in,
width=3.5544in
]{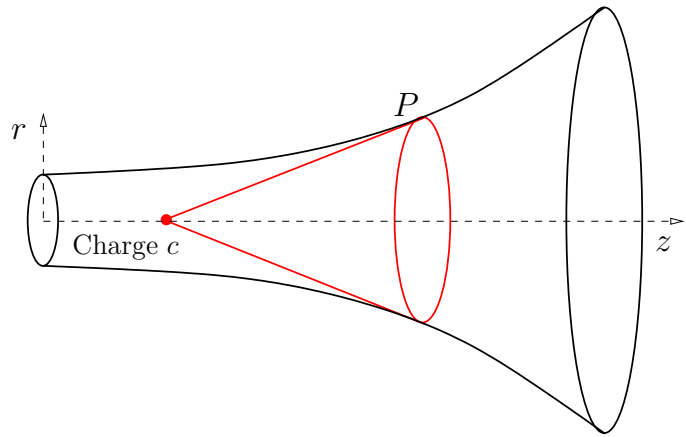}
\end{center}
\caption{{\protect\footnotesize {The "near vortex" approximation.}}}%
\label{conic}%
\end{figure}

To determine the solution we start with an approximation that is valid when
the junction is near the vortex (soon it will be clear what we mean). The
difficult part of the master equations is the partial differential equation
that determines the magnetic field. To overcome this problem we use an
approximation that is valid only when the profile is near to be a cone. To
determine the $B$ field we proceed with the following steps. We take a point
$P=(z,r=f(z))$ on the surface and then we consider the cone that passes
through it and is tangent to the surface (see Figure\ \ref{conic}).
Then we find the magnetic field that is generated by a charge $c$
at the tip of the cone%
\begin{equation}
|\vec{B}|=\frac{c}{d^{2}}=c\left(  \frac{f^{\prime}}{f}\right)  ^{2}\frac
{1}{1+\left(  \frac{rf^{\prime}}{f}\right)  ^{2}}\ . \label{magnetic}%
\end{equation}
where $d$ is the distance from a generic point $(z,r)$ on the red surface to
the tip of the cone. In order to have a constant magnetic flux as we vary $z$,
the charge $c$ must be a function of $z$. The magnetic flux is
\begin{equation}
\Phi_{B}=\int B_{z}=2\pi\int_{0}^{f^{\prime}}dr\,rc\left(  \frac{f^{\prime}%
}{f}\right)  ^{2}\frac{1}{\left(  1+\left(  \frac{rf^{\prime}}{f}\right)
^{2}\right)  ^{3/2}}~.
\end{equation}
Performing the integral we finally obtain:
\begin{equation}
\Phi_{B}=2\pi c\left(  1-\frac{1}{\sqrt{1+{f^{\prime}}^{2}}}\right)  ~.
\label{constant}%
\end{equation}
This expression gives us the function $c(z)$. The magnetic field in the point
$P$ is what we need for the differential equation (\ref{rescaled}). To obtain
it we just take (\ref{magnetic}) evaluated at $r=f(z)$. The function $c(z)$ is
given by (\ref{constant}) where we have also to remember that in our rescaled
unit where the radius of the vortex is $1$ and the magnetic field inside is
$1$, the magnetic flux is $\pi$. Finally the differential equation
(\ref{rescaled}) becomes:
\begin{equation}
-\frac{f^{\prime}{}^{4}}{4f^{4}\left(  1+f^{\prime2}\right)  \left(
\sqrt{1+f^{\prime2}}-1\right)  ^{2}}+\rho\frac{1+f^{\prime2}-~f^{\prime\prime
}f}{f(1+f^{\prime2})^{3/2}}+(1-\rho)=0~. \label{ordinary}%
\end{equation}
As we mentioned before this differential equation is only an approximation. The
procedure we have used to compute the magnetic field is not exact but becomes
reliable when the surface is very well approximated by a cone. This means that
the quantity $ff^{\prime\prime}$ must be very small. Near the vortex the
second derivative $f^{\prime\prime}$ is very small and thus we can use the
ordinary differential equation (\ref{ordinary}) to understand the physical
properties of the junctions near the vortex.

\begin{figure}[!htb]
\begin{center}
\subfigure{\includegraphics[width=.45\linewidth]{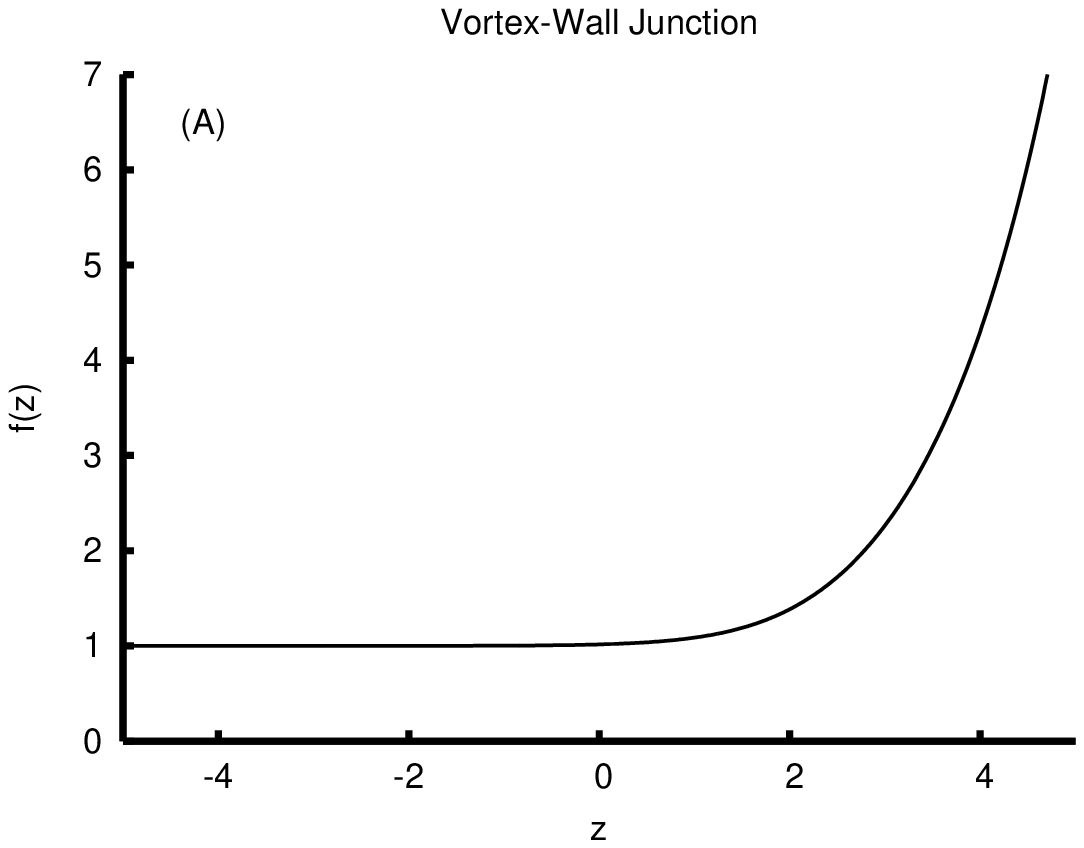}} \qquad
\subfigure{\includegraphics[width=.45\linewidth]{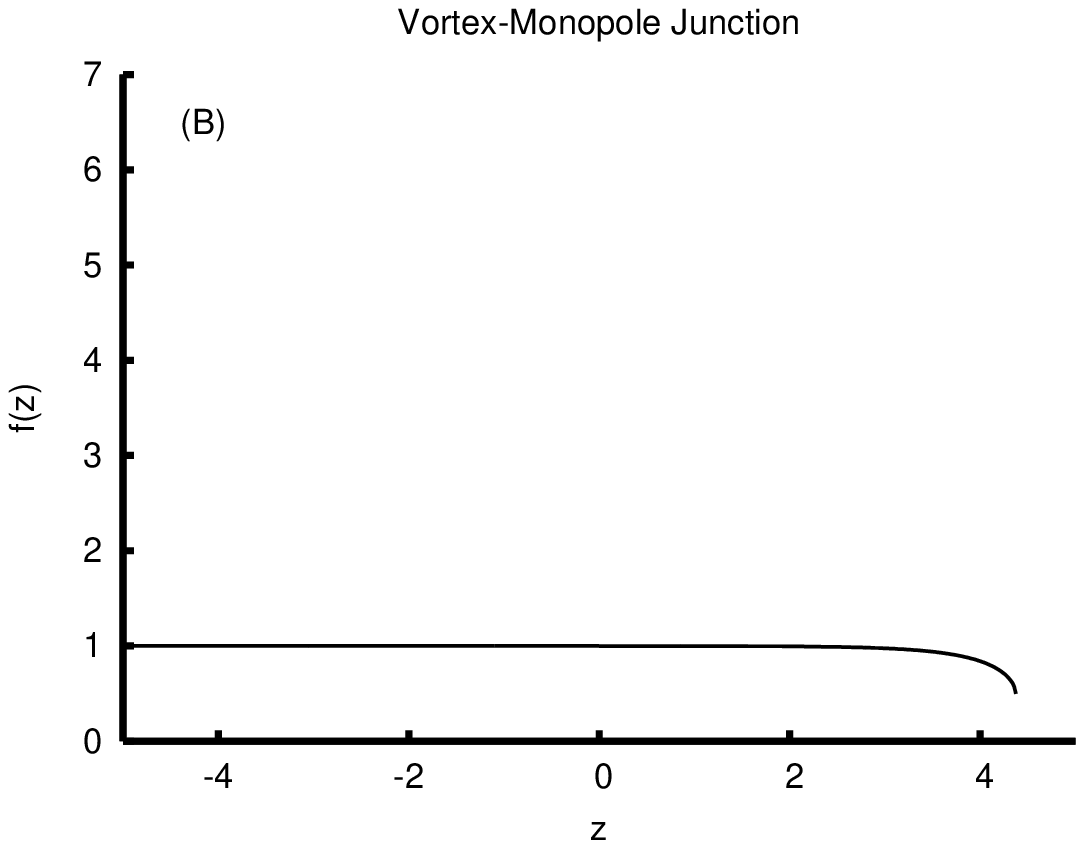}}
\end{center}
\caption{{\protect\footnotesize {The 
orbits.}}}%
\label{homoclin}%
\end{figure}

\begin{figure}[!htb]
\begin{center}
\subfigure{\includegraphics[width=.45\linewidth]{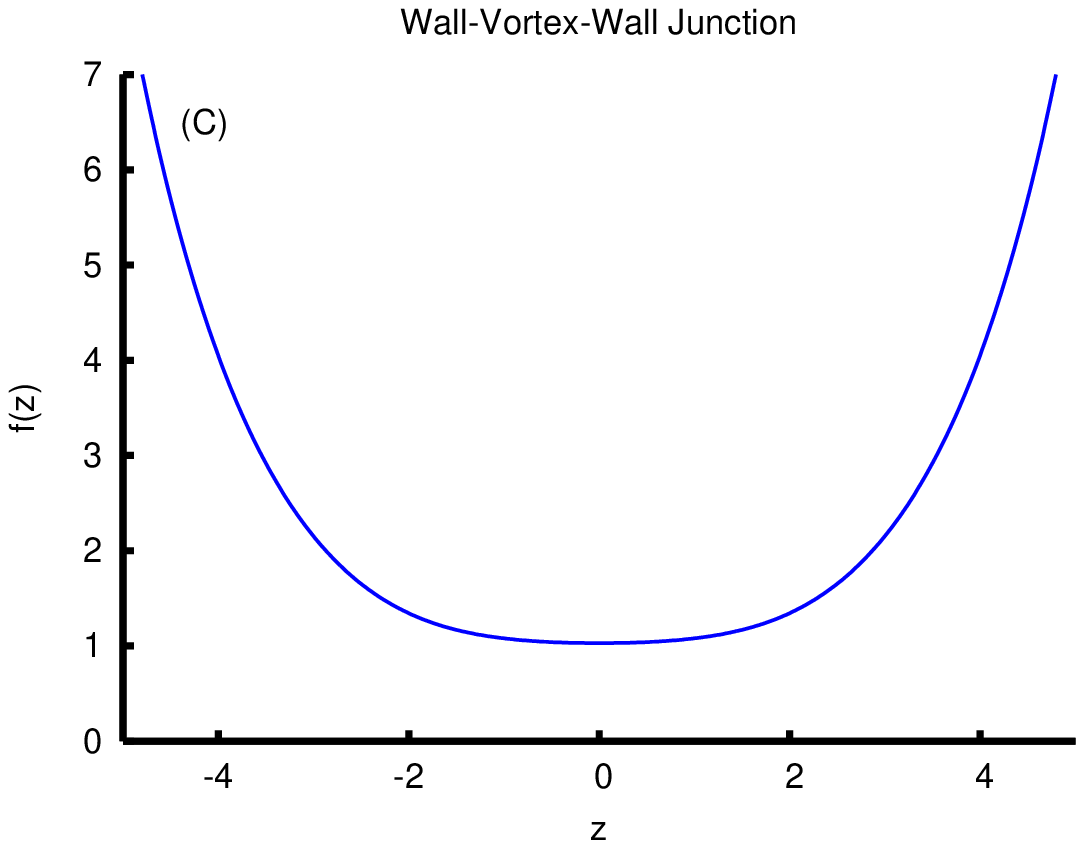}} \qquad
\subfigure{\includegraphics[width=.45\linewidth]{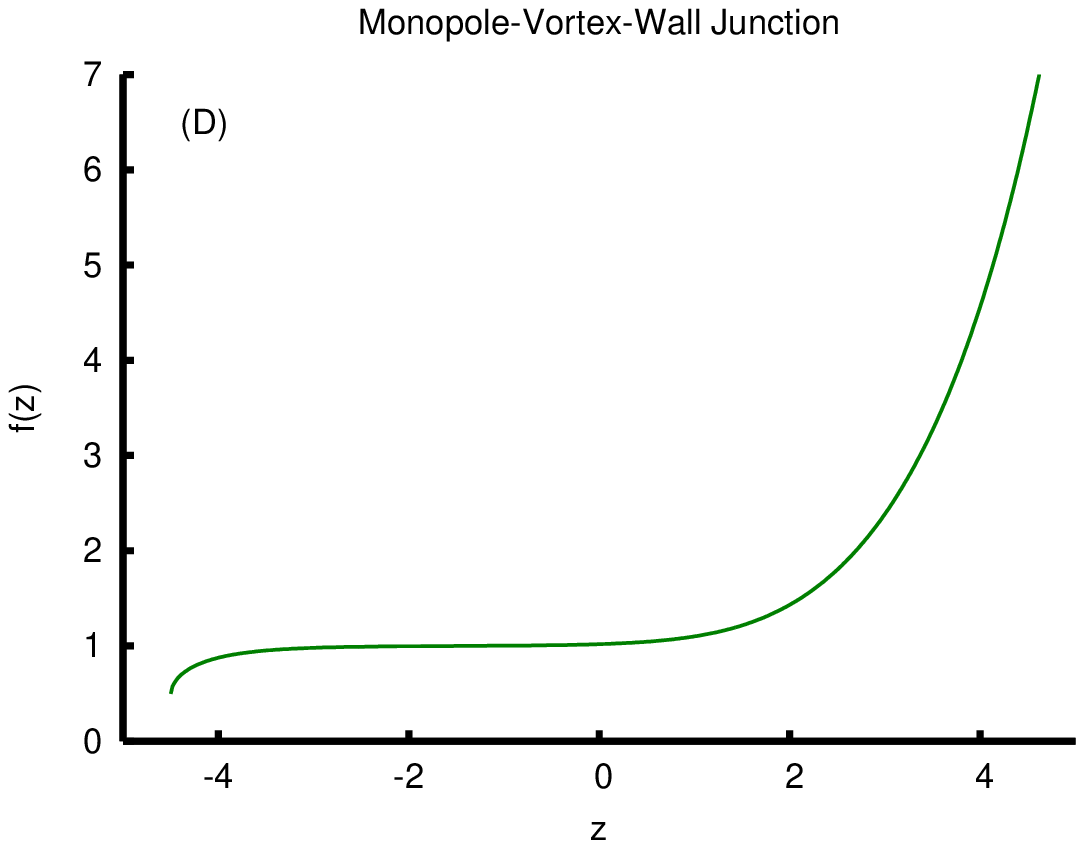}}
\subfigure{\includegraphics[width=.45\linewidth]{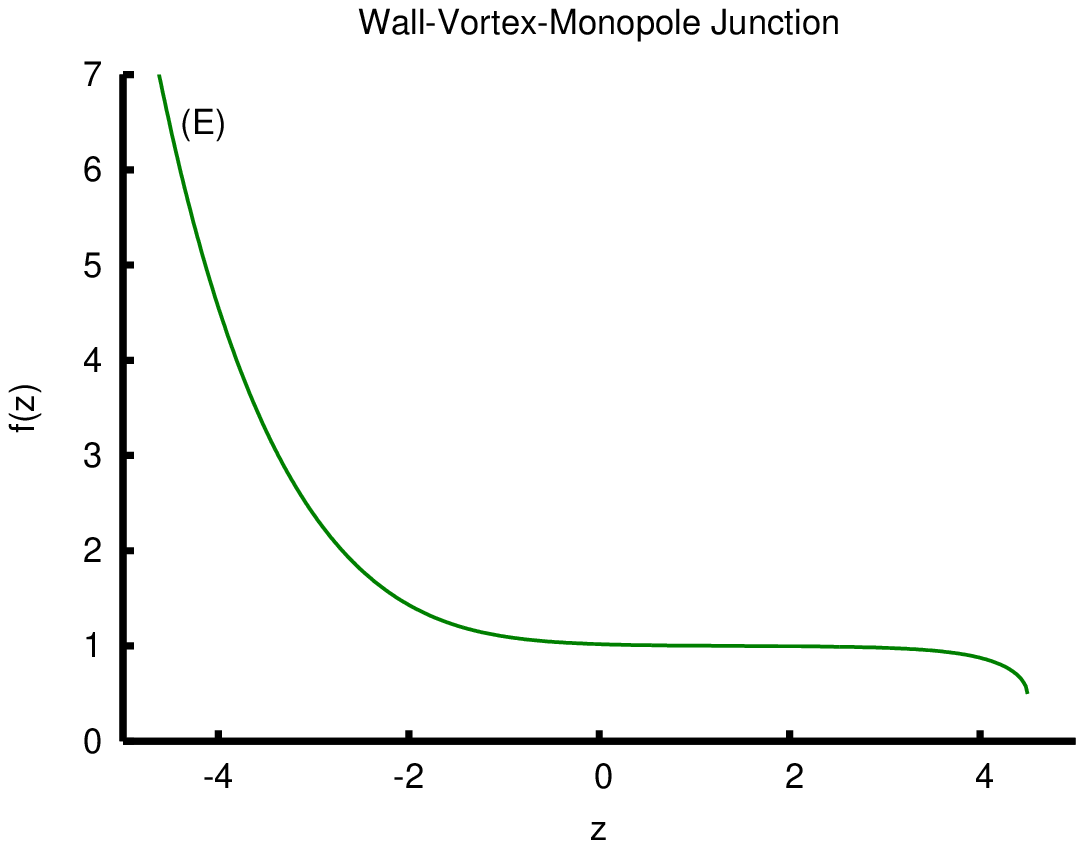}} \qquad
\subfigure{\includegraphics[width=.45\linewidth]{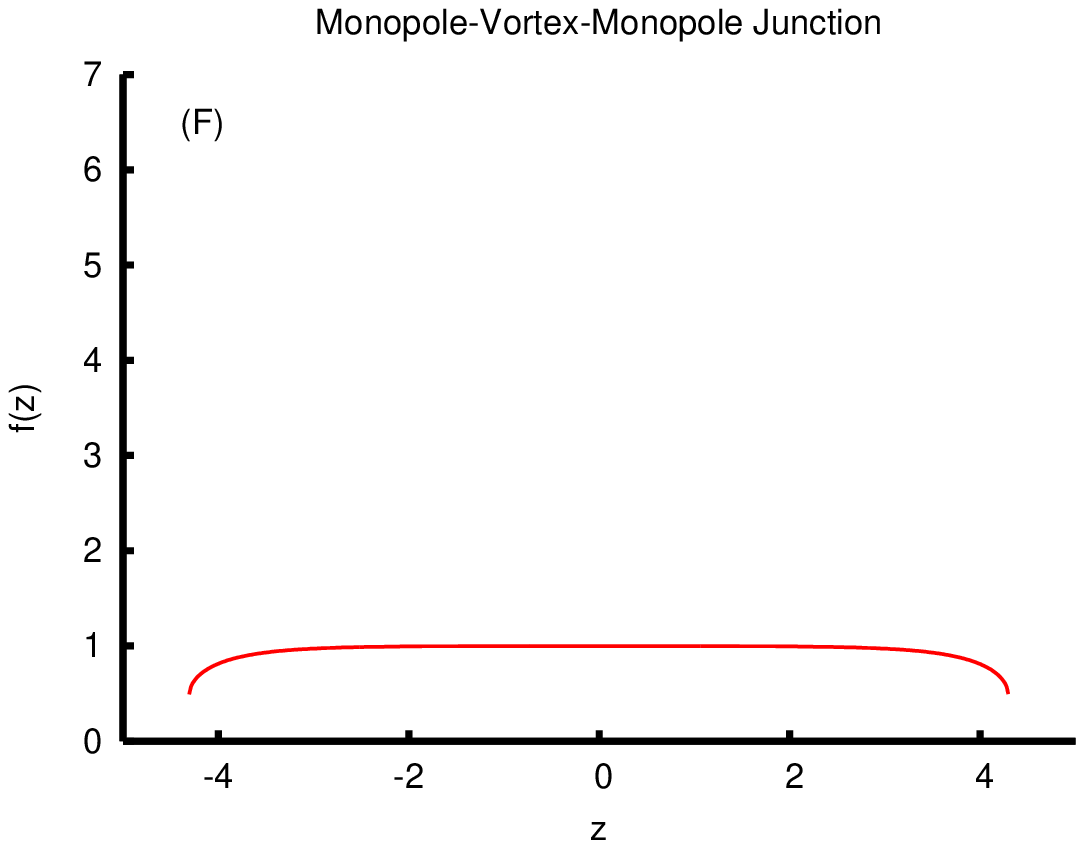}}
\end{center}
\caption{{\protect\footnotesize {More general junctions.}}}%
\label{general}%
\end{figure}

Now we take a look at the solutions to the differential equation.
From now on we use $\rho=1$ that means only tension and zero energy
density. In Figure \ref{homoclin} we have the two fundamental
junctions: (A) is a vortex connected to a domain wall and (B) a
vortex that ends on a point. Since this point must be a source of
magnetic flux, we can call it a monopole. In Figure \ref{general} we
have more general junctions: (C) is a wall-vortex-wall, (D) is a
monopole-vortex-wall, (F) monopole-vortex-monopole.


\begin{figure}[h!tb]
\begin{center}
\includegraphics[
height=3.5405in,
width=5.0548in
]{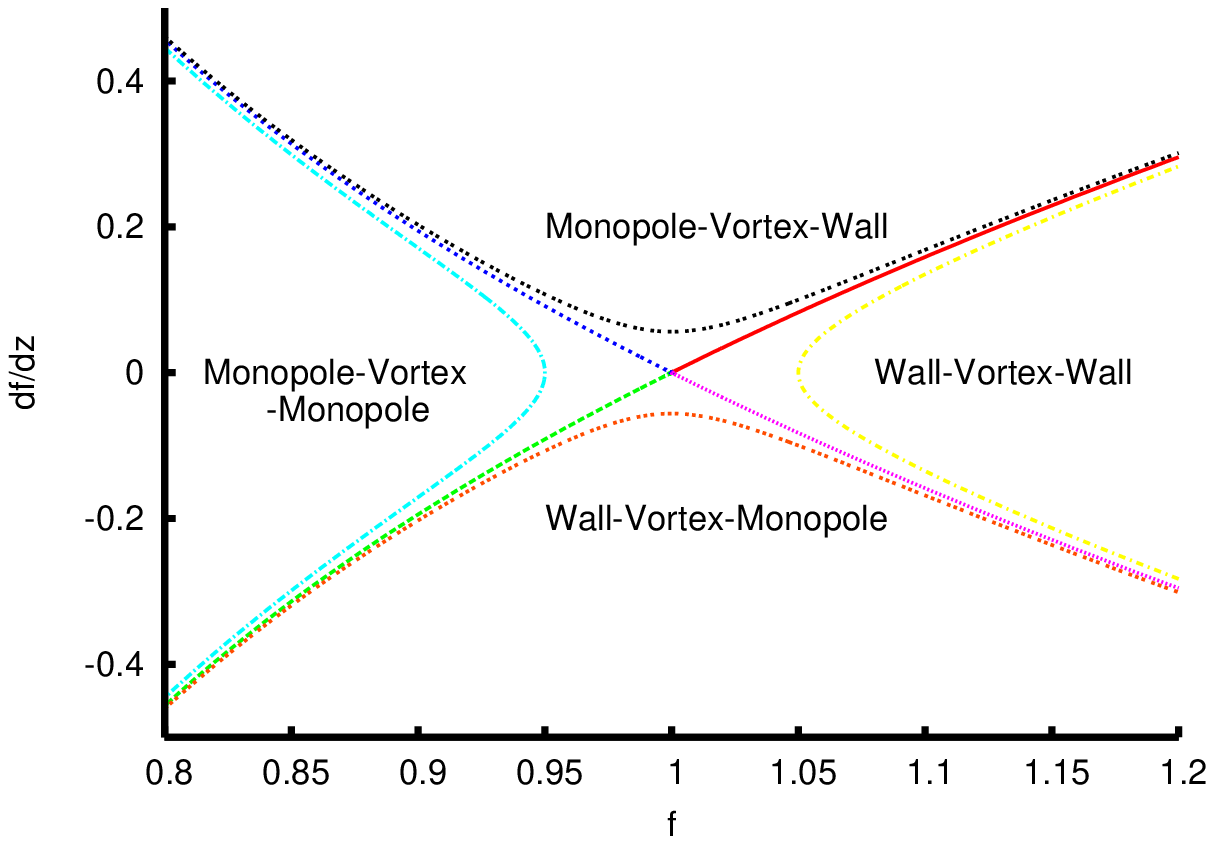}
\end{center}
\caption{{\protect\footnotesize {The phase diagram $(f,f^{\prime})$ around the
stationary point $(1,0)$.}}}%
\label{phase}%
\end{figure}

The overall picture becomes more clear if we consider the differential
equation (\ref{ordinary}) as a \emph{dynamical system}. This means simply that
we rewrite the second order differential equation in the following way%
\begin{equation}
\left\{
\begin{array}
[c]{r@{\hspace{1.5pt}}l}%
f^{\prime}&=g\\
g^{\prime}&=\mathcal{F(}f,g\mathcal{)}%
\end{array}
\right. \ ,  \label{dynamical}%
\end{equation}
where $\mathcal{F(}f,g\mathcal{)}$ is just what remains if we isolate
$f^{\prime\prime}$ in the differential equation (\ref{ordinary}). The
differential equations (\ref{dynamical}) are a simple example of a
dynamical system. We have
a phase space $(f,g)$ and a flow defined on it. The time of the flow is in
this case the $z$ coordinate. The vortex is the fixed point of the
flow $f=1$ and 
$g=0$. To understand the physical behavior around the fixed point we make a
Taylor expansion
\begin{equation}
\left(
\begin{array}
[c]{c}%
f^{\prime}\\
g^{\prime}%
\end{array}
\right)  =\left(
\begin{array}
[c]{cc}%
0 & 1\\
3 & 0
\end{array}
\right)  \left(
\begin{array}
[c]{c}%
f-1\\
g
\end{array}
\right)  +\dots
\end{equation}
The diagonalization leads to
\begin{equation}
\left(  g\pm\sqrt{3}\left(  f-1\right)  \right)  ^{\prime}=\pm\sqrt{3}\left(
g\pm\sqrt{3}\left(  f-1\right)  \right)  +\dots\label{linear}%
\end{equation}
What we read from this is the following. The vortex is a stationary
point of the dynamical system. The linear expansion shows that this
is a \emph{saddle point}. The solutions of Figure \ref{homoclin}
(with the corresponding $z\rightarrow-z$ reflected ones) are the 
orbits that at $z\rightarrow-\infty$
($z\rightarrow\infty$) go into the vortex. The 
orbits
around the vortex are the two lines (\ref{linear}) $g=\pm
\sqrt{3}\left(  f-1\right)  $. If we draw the solutions in a phase plot
we obtain Figure \ref{phase}. 

\subsection{Flux tube, domain wall and monopole ($\rho=1$)\label{rhoone}}

The first junction we want to study is the flux tube that ends on a domain
wall. In this case we have to take $\rho=1$ so the Higgs phase and the
Coulomb phase are both true vacua of the theory.

\begin{figure}[h!tb]
\begin{center}
\includegraphics[
height=2.8392in,
width=4.5662in
]{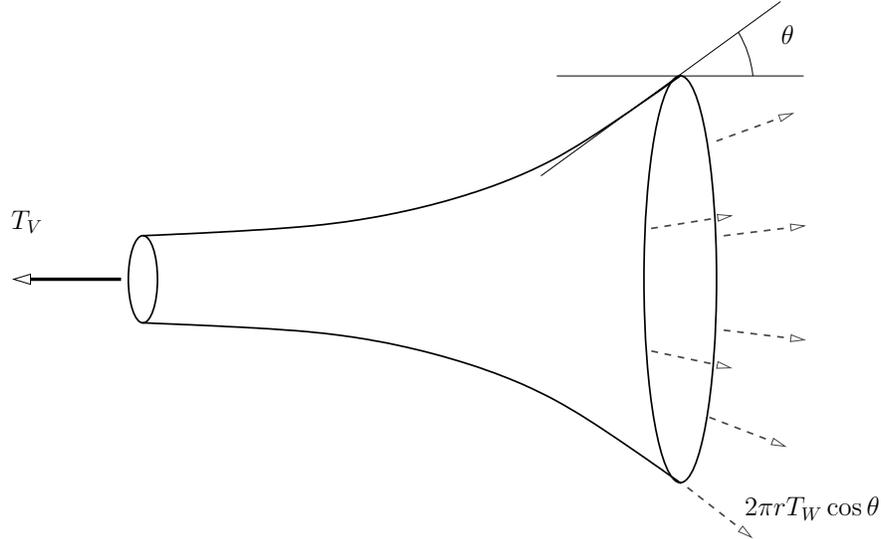}
\end{center}
\caption{{\protect\footnotesize {Asymptotic behavior of the logarithmical
bending.}}}%
\label{tension}%
\end{figure}

First we want determine the coefficient of the logarithmic
deformation of the wall when a vortex is attached to it. What we are
going to describe is summarized in Figure \ref{tension}.
The piece of junction in the figure has two forces that acts on
it. The first is the vortex that pulls it down with a tension $T_{V}$. The
second is the wall that pulls it up. These two forces must be equal:
\begin{equation}
T_{\mathrm{V}}=2\pi rT_{\mathrm{W}}\frac{1}{\sqrt{1+{\tan{\theta}}^{2}}}\ ,
\end{equation}
where $\tan{\theta}=\mathrm{d}r/\mathrm{d}z$. At the end we obtain a
differential equation for the profile $r(z)$:
\begin{equation}
{r^{\prime}(z)}^{2}-(2\pi T_{\mathrm{W}}/T_{\mathrm{V}})^{2}{r(z)}^{2}+1=0\ .
\label{large}%
\end{equation}
This equation can be trusted only at large $r$. In fact in making the
balance of 
the forces we have not considered the magnetic field inside the wall. For
large $r$ its contribution is negligible since $B$ fall offs as
$1/r^{2}$ which implies that the energy density $B^{2}r^{2}$ falls of
as $1/r^{2}$. Anyway (\ref{large}) 
gives the correct coefficient of the logarithmic bending at large $r$:
\begin{equation}
z\simeq\frac{T_{\mathrm{V}}}{2\pi T_{\mathrm{W}}}\log{r}+\cos t\ .
\end{equation}
Since $\frac{T_{\mathrm{V}}}{2\pi T_{\mathrm{W}}}=\frac{3}{2}R_{\mathrm{V}}$,
and we are choosing units in which $R_{\mathrm{V}}=1$, this means that the
vortex-wall junction asymptotically goes like%
\begin{equation}
r\propto e^{\frac{2}{3}z}~. \label{exponential}%
\end{equation}

\begin{figure}[h!tb]
\begin{center}
\subfigure{\includegraphics[width=.48\linewidth]{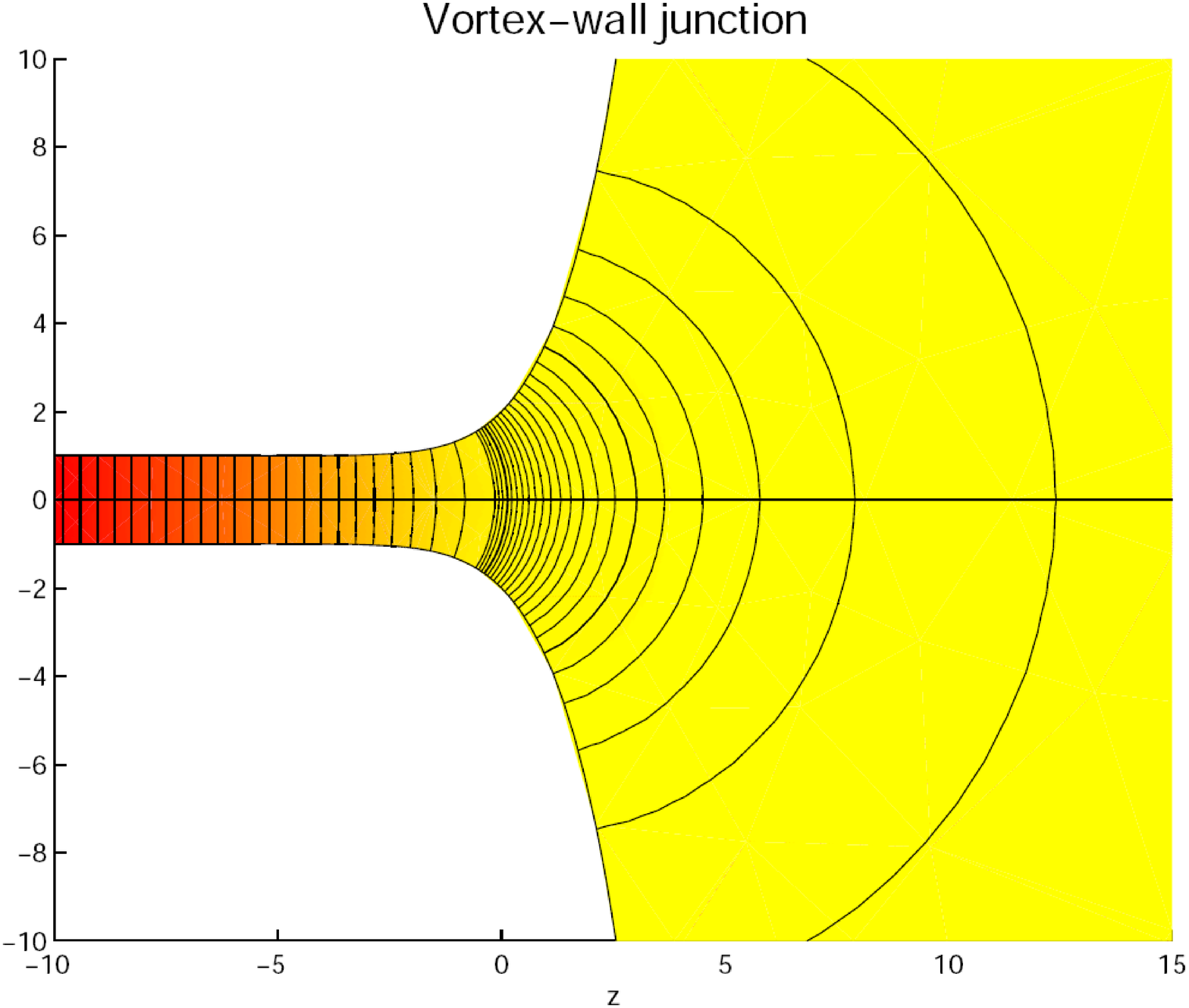}}
\qquad\subfigure{\includegraphics[width=.45\linewidth]{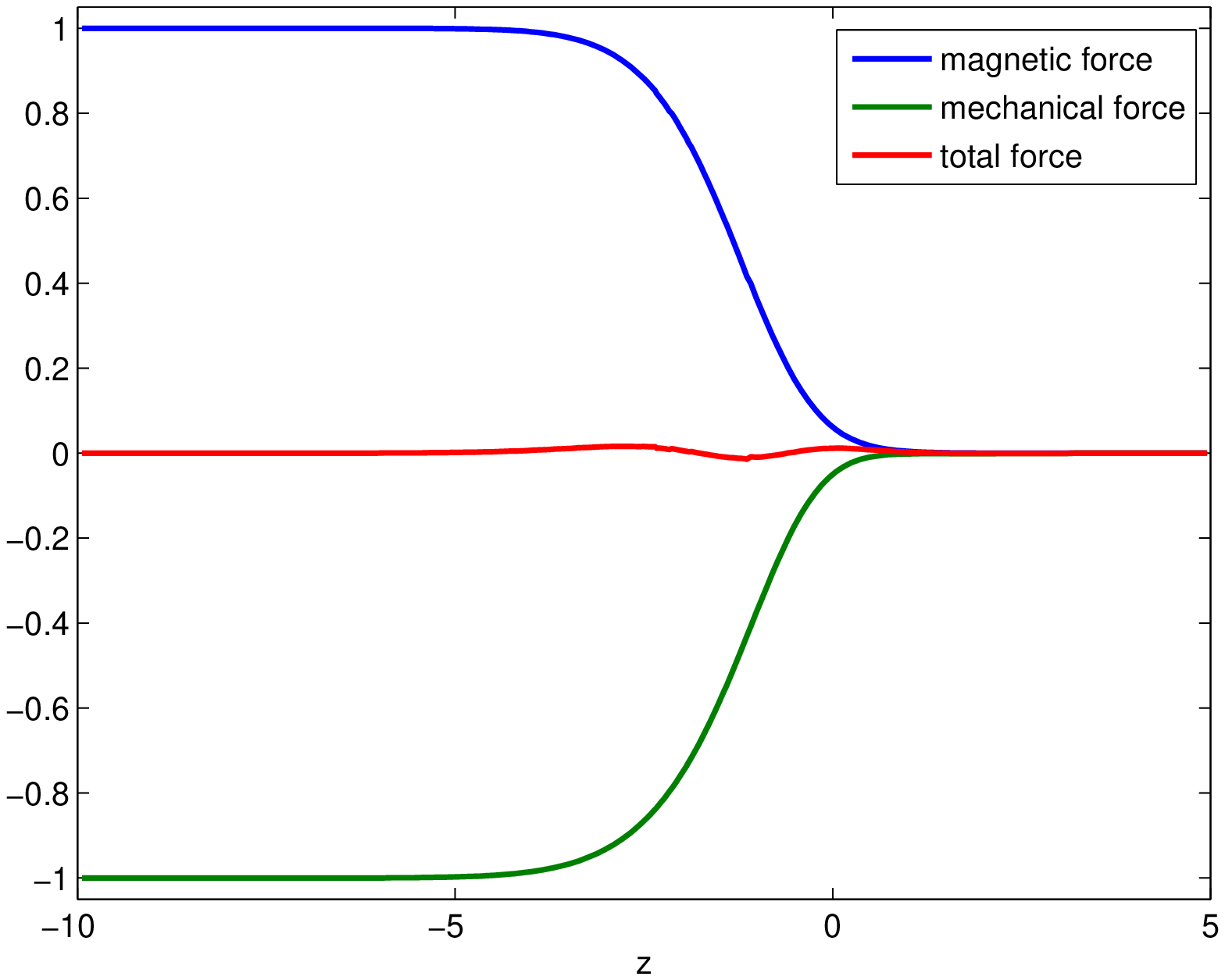}}
\end{center}
\caption{{\protect\footnotesize {Vortex-wall junction and its force plot
($\rho=1$).}}}%
\label{vwjunction}%
\end{figure}

\begin{figure}[h!tb]
\begin{center}
\subfigure{\includegraphics[width=.48\linewidth]{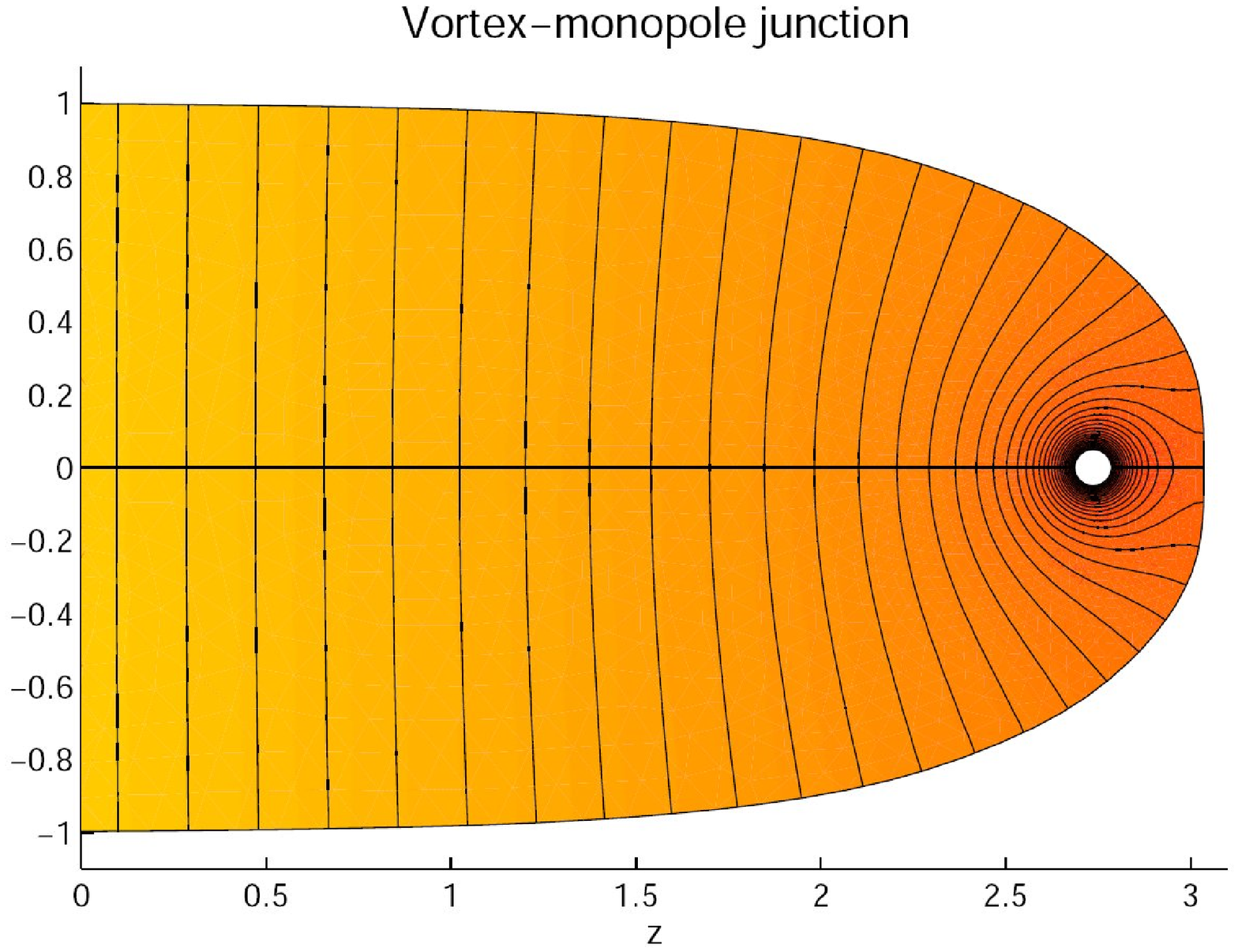}}
\qquad \subfigure{
\includegraphics[width=.40\linewidth]{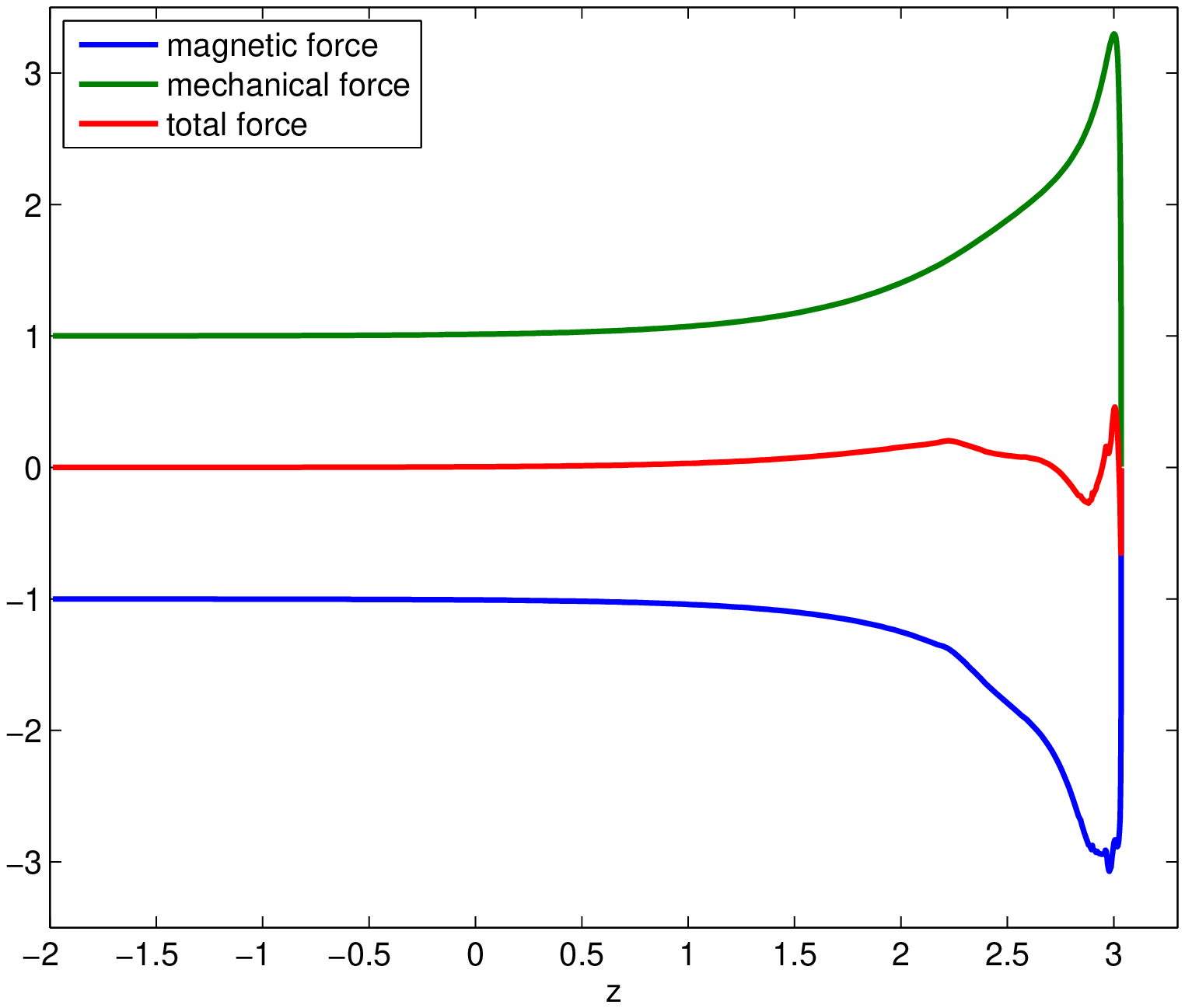}}
\end{center}
\caption{{\protect\footnotesize {Vortex-monopole junction and its force plot
($\rho=1$).}}}%
\label{vmjunction}%
\end{figure}

\begin{figure}[h!tb]
\begin{center}
\includegraphics[width=.42\linewidth]{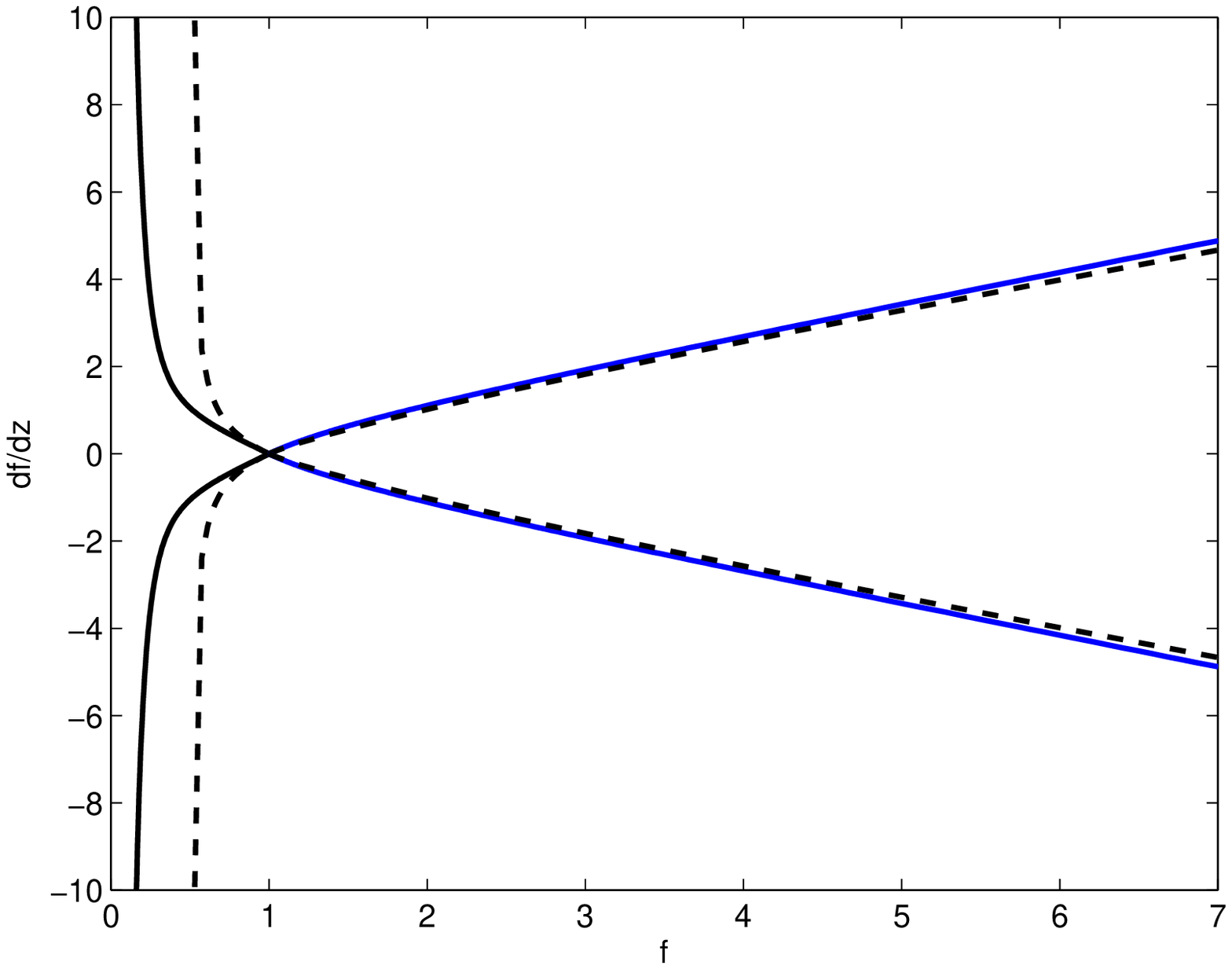}
\end{center}
\caption{{\protect\footnotesize Phase diagram for $\rho=1$. The
dashed lines correspond the near vortex approximation given in Figure \ref{phase}.}}%
\label{phaseuno}%
\end{figure}

Now it is time to finally compute the vortex-wall junction. The
strategy we use is the following. Eqs. (\ref{linear}) and
(\ref{exponential}) give us the asymptotic behavior of the profile
function at $z\rightarrow-\infty$ (near the vortex) and
$z\rightarrow+\infty$ (near the wall). We construct a trial function
with the desired asymptotic behaviors and with a certain number of
free parameters. These parameters will be adjusted by our numerical
code in order to approximate in the best possible way the
\textquotedblleft real\textquotedblright\ vortex-wall solution. The
more the free parameters are the bigger is the space of profile
functions that our trial function can span. For any given choice of
the parameters we have a certain profile function and we can thus
solve the Laplace equation (\ref{uno}) and the mechanical forces. To
solve the Laplace equation we use a finite element method routine.
Once we have the mechanical force and the magnetic force for a given
profile we can compute the total force that is just
the sum of the two. For the \textquotedblleft real\textquotedblright%
\ vortex-wall junction the two forces must be exactly equal but
opposite in direction in every point on the profile (\ref{due}). For
our trial function we thus compute the norm of the total force (the
integral of the total force squared) and then minimize it. The
minimization gives, within the space spanned by our trial functions,
the best approximation to the real junction. For the vortex-wall
junction the result of the computation is given in Figure
\ref{vwjunction}. We show both the junction and the force diagram.
We have used the same strategy to compute the vortex-monopole
junction. The result is given in Figure
\ref{vmjunction}.
We can finally plot the global phase diagram in Figure
\ref{phaseuno}.

\subsection{Flux tube, domain wall and monopole ($\rho<1$)\label{rholess}}

Now we want to study the same junctions in the case $\rho<1$. We can expand
the \textquotedblleft near vortex\textquotedblright%
\ approximation\ (\ref{ordinary}) around the stationary point $(1,0)$ and we
obtain%
\begin{equation}
\left(
\begin{array}
[c]{c}%
f^{\prime}\\
g^{\prime}%
\end{array}
\right)  =\left(
\begin{array}
[c]{cc}%
0 & 1\\
\frac{4-\rho}{\rho} & 0
\end{array}
\right)  \left(
\begin{array}
[c]{c}%
f-1\\
g
\end{array}
\right)  +\dots
\end{equation}
The diagonalization leads to
\begin{equation}
\left(  g\pm\sqrt{\frac{4-\rho}{\rho}}\left(  f-1\right)  \right)  ^{\prime
}=\pm\sqrt{\frac{4-\rho}{\rho}}\left(  g\pm\sqrt{\frac{4-\rho}{\rho}}\left(
f-1\right)  \right)  +\dots
\end{equation}
And so the 
orbits around the vortex are the two lines
$g=\pm\sqrt{\frac{4-\rho}{\rho}}\left(  f-1\right)  $.

\begin{figure}[h!tb]
\begin{center}
\subfigure{\includegraphics[width=.50\linewidth]{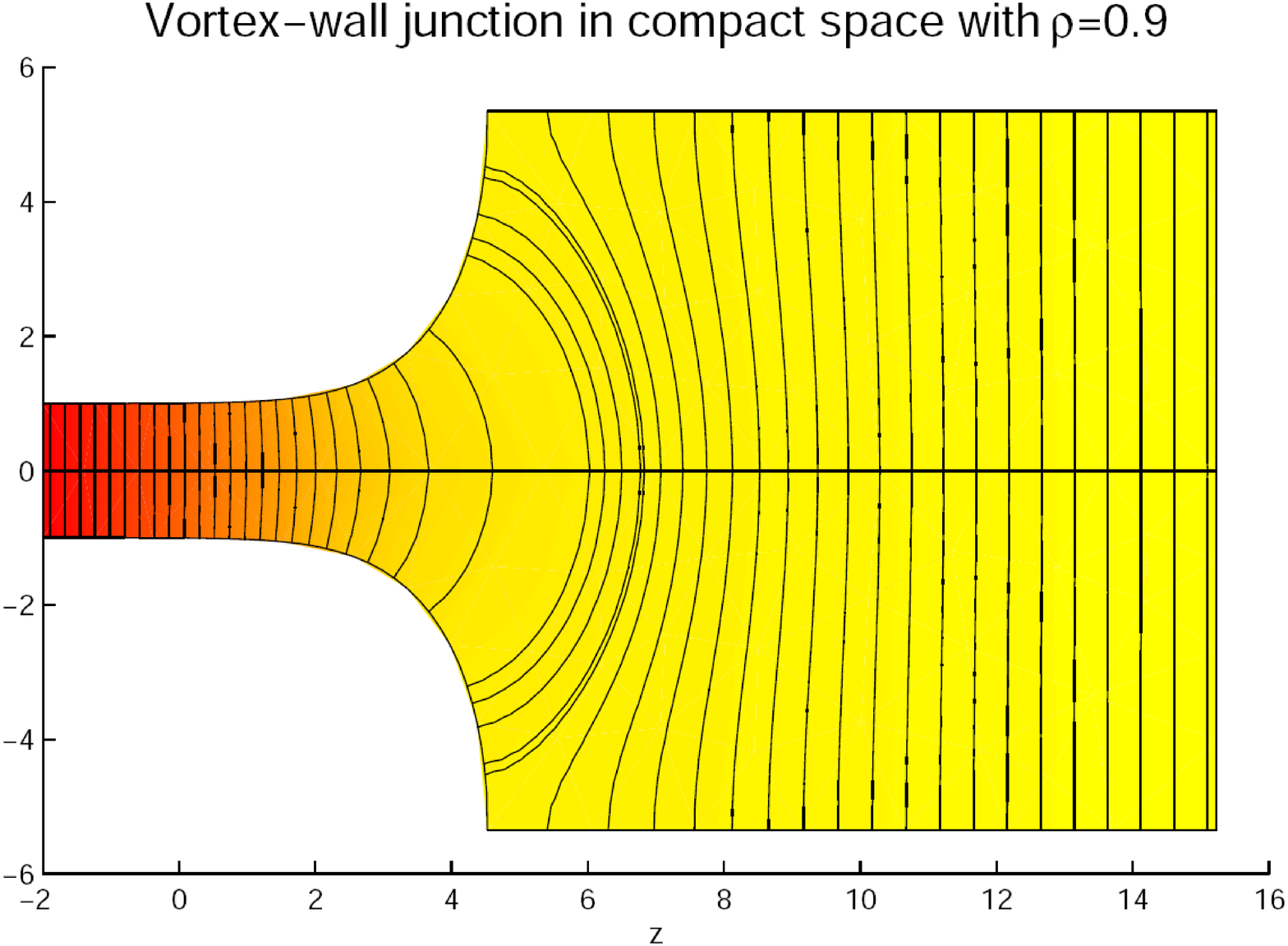}}\qquad
\subfigure{\includegraphics[width=.36\linewidth]{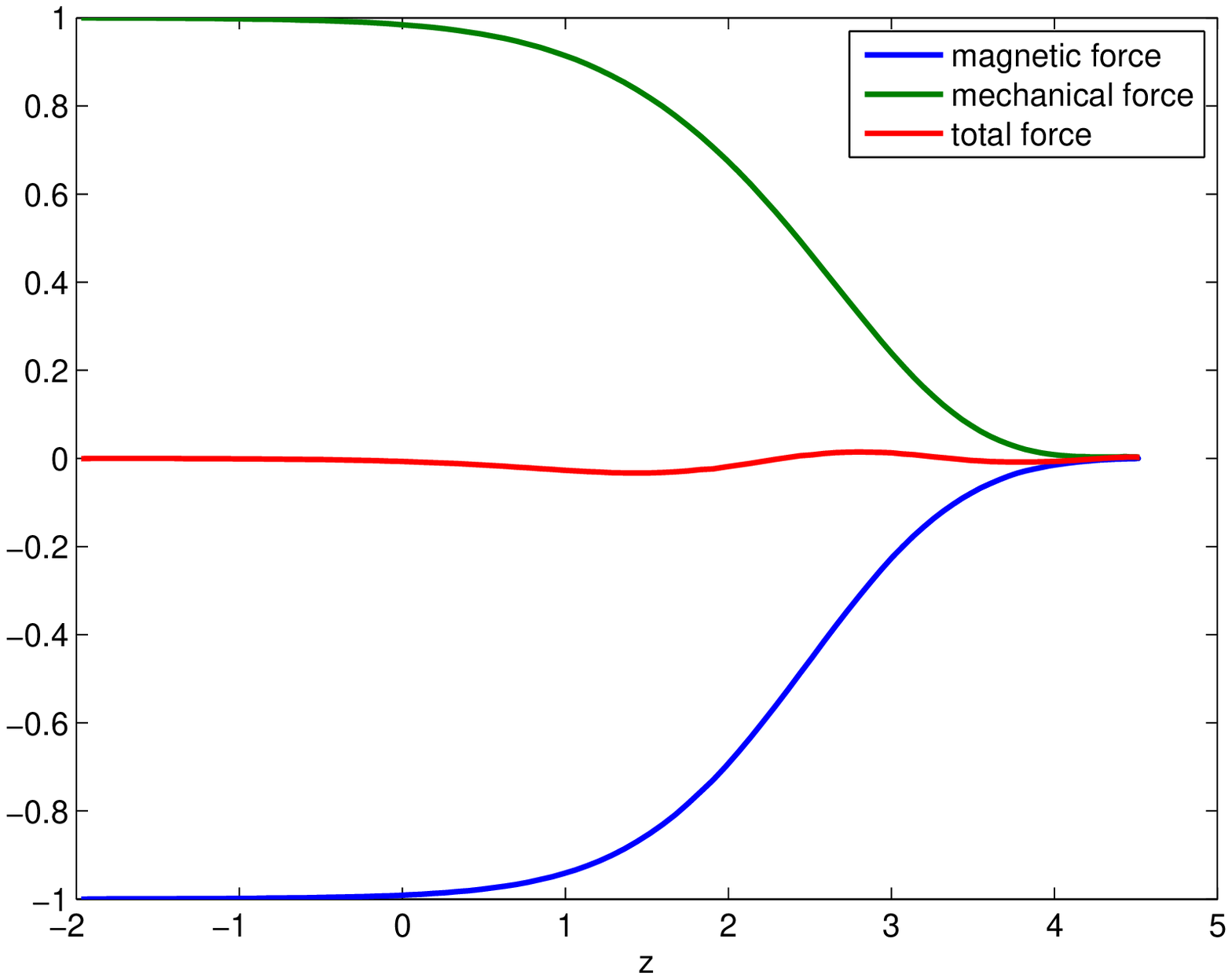}}
\end{center}
\caption{{\protect\footnotesize {Vortex-wall junction and its force plot
($\rho=0.9$).}}}%
\label{vwjunctionless}%
\end{figure}

The first junction we want to study is the vortex-wall. In the case
$\rho<1$ the energy density of the Coulomb vacuum is not zero and
this means that a stable domain wall does not exist. On the other
hand we have a beautiful solution for the vortex-wall junction in
the case $\rho<1$ in Figure \ref{vwjunctionless}. What happens to this
solution when $\rho$ is decreased? The fact is that a vortex-wall
junction can exist if the two dimensional space orthogonal to the
vortex is compactified. If we compactify the two dimensional space
on a circle then its radius $R_{\mathrm{Max}}$ is given by
\begin{equation}
T_{\mathrm{V}}=\frac{\Phi_{B}^{2}}{2\pi R_{\mathrm{V}}^{2}}+T_{\mathrm{W}}2\pi
R_{\mathrm{V}}+\varepsilon_{0}\pi R_{\mathrm{V}}^{2}\ =\frac{\Phi_{B}^{2}%
}{2\pi R_{\mathrm{Max}}^{2}}+\varepsilon_{0}\pi R_{\mathrm{Max}}^{2}~.
\end{equation}
The previous equation is simply the balance of forces. The tension
of the vortex, that is equal to the sum of the magnetic field, wall and
bulk energies, must be equal to the tension in the Coulomb vacuum that
is given only by the magnetic field and the energy density. With our units ($\Phi_{B}%
=\pi,T_{\mathrm{W}}=\frac{\rho}{2},\varepsilon_{0}=\frac{1-\rho}{2}$) the
radius $R_{\mathrm{Max}}$ is determined by the following equation%
\begin{equation}
\pi\left(  1+\frac{\rho}{2}\right)  =\frac{\pi}{2R_{\mathrm{Max}}^{2}}%
+\frac{1-\rho}{2}\pi R_{\mathrm{Max}}^{2}~.
\end{equation}
For the vortex-wall junction the result of the computation is given
in Figure \ref{vwjunctionless}.

\begin{figure}[h!tb]
\begin{center}
\subfigure{\includegraphics[width=.45\linewidth]{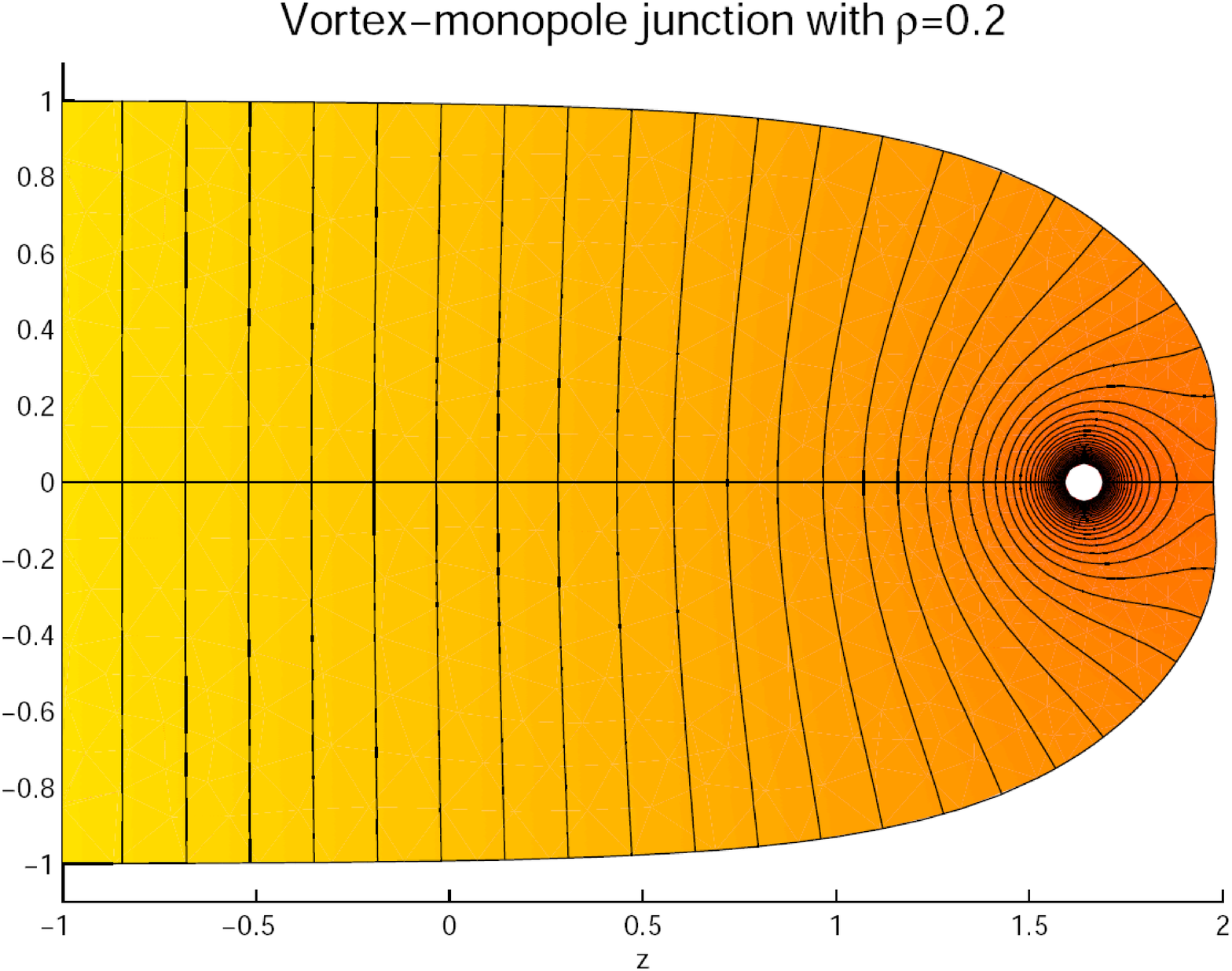}}\qquad\subfigure{
\includegraphics[width=.46\linewidth]{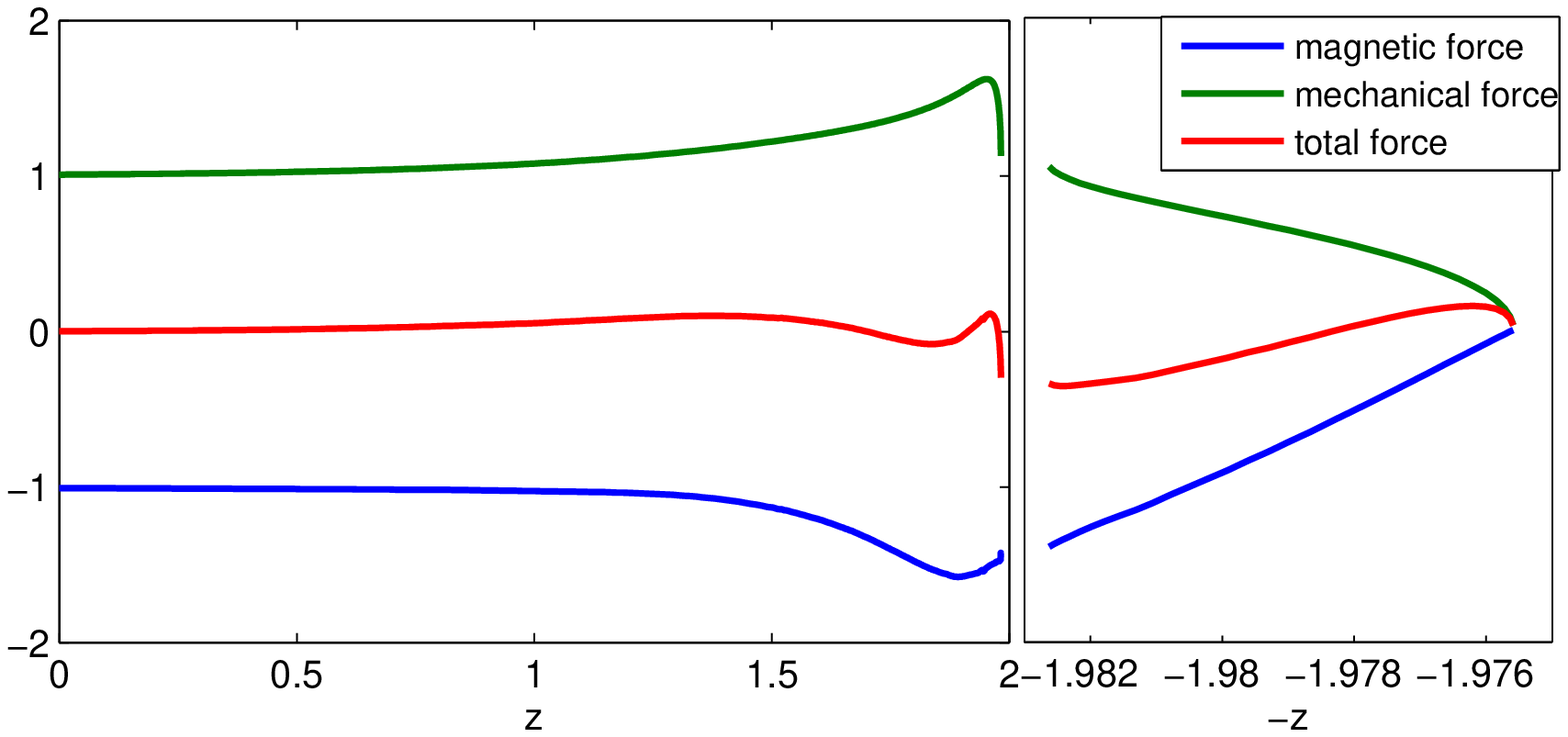}}
\end{center}
\caption{{\protect\footnotesize {Vortex-monopole junction and its force plot
($\rho=0.2$).}}}%
\label{vmjunctionless}%
\end{figure}

\begin{figure}[h!tb]
\begin{center}
\includegraphics[width=.42\linewidth]{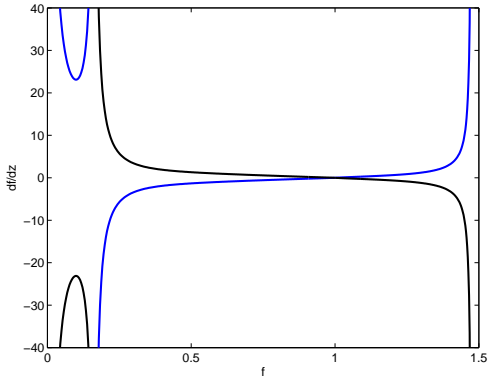}
\end{center}
\caption{{\protect\footnotesize {Phase diagram for $\rho=0.2$.}}}%
\label{phasedue}%
\end{figure}

The last junction is the vortex-monopole junction for the case
$\rho<1$. This junction is the most difficult to compute and is also
the one where we have reached least precision. The difficult part
comes from the shape of the wall vortex surface around the point
where it intersects the axial line $r=0$. Due to symmetry reasons
the magnetic field in this point must be zero so the total
mechanical force must be zero. In the case $\rho=1$ the Coulomb
energy density $\varepsilon_{0}$ is zero and so also the curvature
of the wall vortex surface must be zero (see Figure
\ref{vmjunction}). When the Coulomb energy density $\varepsilon_{0}$
is not zero and it exerts a force that tends to retreat the surface.
The two radia of curvature are equal, directed outwards and with
modulus given by
\begin{equation}
2\frac{T_{\mathrm{W}}}{R_{\mathrm{Curv}}}=\varepsilon_{0} \ , %
\end{equation}
that in our units means%
\begin{equation}
R_{\mathrm{Curv}}=\frac{2\rho}{1-\rho}~.
\end{equation}
For the vortex-monopole junction the result is given in Figure
\ref{vmjunctionless}. 
We also plot the global phase for $\rho=0.2$ in Figure
\ref{phasedue}.

\section{Conclusions and Discussions \label{conclusion}}

We conclude the paper with a discussion on possible generalizations and
physical applications of the junctions of large $n$ vortices.

\subsection{Soliton junctions and dynamical systems\textit{~~~}}

One interesting outcome of this paper is the relation between soliton
junctions and a dynamical system (see \cite{dynamicaluno} for an introduction
to the subject). Considering $z$ as the time of the dynamical system, the flux
tube is a stationary point of the differential equation. The linear expansion
around it shows that this is a saddle point where two lines of 
orbits intersect. Now think about the junction with a vortex in the middle,
for example the wall-vortex-wall or the monopole-vortex-monopole. We can have
a lot of this kind of junctions since the length of the vortex in the middle
can be fixed at pleasure. From the dynamical system point of view this is very
simple to understand. Since the vortex is a saddle point we can have orbits
that goes arbitrarily near to the vortex\ and then escape.

\ In the analysis of Subsection \ref{nearvortexapproximation} we have a very
simple dynamical system. The phase space is two dimensional and consist of the
profile $f(z)$ and its derivative $f^{\prime}(z)$. Now suppose we are dealing
with a \emph{non}-large $n$ vortex. The phase space will be an \emph{infinite}
dimensional space This is the space of the field configurations $q(r,z)$ and
$A_{i}(r,z)$ and their derivatives. Even in this infinite dimensional space we
can think of the vortex as a stationary saddle point of the dynamical system.
The fact that the vortex is a saddle point is just a consequence of the
$z\rightarrow-z$ reflection symmetry. If we make a linearization around a
stationary point and we have a subspace $H^{(-)}$ of 
orbits that go
to the vortex solution at $z\rightarrow-\infty$, we also have a mirror
symmetric subspace $H^{(+)}$ of 
orbits that go to the vortex
solution at $z\rightarrow+\infty$.

The theory of dynamical systems is a very vast and evolutes branches of
mathematics. What we have used in this paper is just a tiny bit of it. A lot
of interesting phenomena arise in the study of the global properties of the
dynamical systems. It would be interesting to adopt this point of view in the
study of soliton junctions in more sophisticated theories.

\subsection{Web of flux tubes\textit{~~}}

\textit{~}It is possible to relax the condition of cylindrical symmetry and
look for generic stationary configurations of the wall vortex surface.
Consider first the case of the junction between three flux tubes. A vortex
that carries $n$ units of flux can be split into two vortices with fluxes
$n_{1}$ and $n_{2}$ where $n=n_{1}+n_{2}$. As long as $\rho$ is different from
zero, the vortices are of type I and this means that the tensions satisfy the
inequality $T_{1}+T_{2}>T$. A non-trivial junction between the three vortices
is thus possible and the angles are uniquely determined by the tensions. If
$\rho=0$ the angle between the two smaller vortices is zero and so the
junction is trivial. It is nevertheless possible to obtain a four-vortex
junction. \ With these basic junctions we can construct a web of flux tubes in
three dimensions.

The master equations (\ref{uno}) and (\ref{due}) can be generalized relaxing
the condition of cylindrical symmetry. We have to take a map from a generic
punctured oriented two dimensional surface to the three dimensional space. The
punctures on the surface are mapped to the flux tubes at infinity and every
handle correspond to an additional flux tube in the web. The master equation
is now a system of two partial differential equations. The first one is just
the Laplace equation (\ref{uno}) for magnetic scalar potential inside the
tubes (with the correct boundary conditions) and the other one is the balance
of forces%

\begin{equation}
-\left.  \frac{B^{2}}{2}\right\vert _{\mathrm{wall}}+\frac{T_{\mathrm{W}}%
}{R_{1}}+\frac{T_{\mathrm{W}}}{R_{2}}+\varepsilon_{0}=0~,
\end{equation}
where now $R_{1}$ and $R_{2}$ are the two radia of curvature of the surface.

\subsection{Confining strings}

A possible physical application of large $n$ vortices is in the context of
confining strings in large $N$ gauge theories (see \cite{Greensite:2003bk} and
\cite{Shifkstrings} for reviews). We briefly review the hypothesis pointed out
in \cite{wallvortexdue} which is the following. Consider an $SU(N)$ pure gauge
theory and denote by $k$-string the string that confines two heavy probes
quark and anti-quark in the $k$-index antisymmetric representation. We then
consider the \emph{saturation limit }that is the limit where we send both $k$
and $N$ to infinity keeping fixed the ratio $x=k/N$. The large $N$ limit is as
usual accompanied by the 't Hooft rescaling of the coupling constant
$g^{2}=\widetilde{g}^{2}N$. A useful physical quantity is the ratio of string
tensions divided by $N$%
\begin{equation}
\mathcal{R}(x,N)=\frac{1}{N}\frac{T(k,N)}{T(1,N)}~.
\end{equation}
$\mathcal{R}(x,N)$ is the a quantity with a smooth saturation limit.
Now comes an assumption. Inside the ${\EuFrak {su}}(N)$ Lie algebra there is a
particular generator, that up to gauge invariance is unique, that
exponentiated passes trough all the elements of the center of the
gauge group. We assume that the $k$-strings are a sort of \emph{dual
vortices} of this $U(1)$. This is of course a non-proved assumption
but the nice thing, as we are going to see, is that it has a very
clear signal that can be tested with lattice computations
\cite{Lucini:2004my}. The fact is that the string tension must now
satisfy two constraints. The first is that in the \emph{free string
limit} ($k$ fixed while $N$ goes to infinity) the tension must be
linear in $k$ plus subleading corrections. On the other hand, based
on the assumption we just made, the tension for the dual $U(1)$
vortex must also be linear when $k$ is large.
The only reasonable way to
combine these two limits is that $\mathcal{R}(x,N)$ in the
saturation limit is the triangular function plus subleading
corrections.
\begin{equation}
\mathcal{R}(x,N)=\min{(x,1-x)+}\mathcal{O(}1/N\mathcal{)}~.
\end{equation}
If this turns out to be correct, the vortex-monopole junction studied in the
present paper will be a good description of the $k$-string quark junction and
will maybe be visible in lattice computations with a large number of colors.

\section{Acknowledgments}

We thank K.~Konishi for useful comments and for the precious collaboration to
the initial stages of the work. We thank also D.Tong for a useful discussion.
The work of S.B. supported by the Marie Curie Excellence Grant under contract MEXT-CT-2004-013510.

\end{document}